\documentclass[letter]{aa} 

\usepackage{graphicx}
\usepackage{txfonts}
\usepackage{color}
\usepackage{xspace}
\usepackage{siunitx}
\usepackage{amsmath}
\usepackage{amssymb}
\usepackage{xcolor}
\usepackage{dashbox}
\usepackage{framed}
\usepackage{lipsum}

\usepackage{hyperref}

\hypersetup{
    colorlinks=true,                                                
        breaklinks=true,
        linkcolor=blue,                                                     
    citecolor=blue,                                                         
        filecolor=blue,                                                     
        urlcolor=blue,
        unicode=false,                                                      
        pdftoolbar=true,                                                    
        pdfmenubar=true,                                                    
        pdffitwindow=false,                                                 
        pdfstartview={Fit},                                                 
        pdftitle={Extended stellar systems in the solar neighborhood},  
        pdfauthor={Stefan Meingast},                        
        pdfsubject={},                                          
        pdfcreator={Stefan Meingast},                       
        pdfkeywords={}, 
        pdfnewwindow=true,                                                  
        pdfdisplaydoctitle=true                                     
}

\makeatletter
\renewcommand*\aa@pageof{, page \thepage{} of \pageref*{LastPage}}
\makeatother

\definecolor{stefan}{rgb}{0.86, 0.08, 0.24}

\definecolor{joao}{rgb}{0.01, 0.75, 0.24}

\newcommand{\code}{\texttt}

\begin{document}

\title{Extended stellar systems in the solar neighborhood\thanks{We acknowledge the simultaneous publication by \citet{roeser18}, who also used Gaia DR2 to present evidence for the existence of the Hyades's tidal tails. The authors of both publications did not know about each other's work and arrived at their conclusions independently.
}}
\subtitle{I. The tidal tails of the Hyades}


\author{Stefan Meingast\inst{1}
                \and Jo\~ao Alves\inst{1,2,3}
        }
            
\institute{Department of Astrophysics, University of Vienna, T\"urkenschanzstrasse 17, 1180 Wien, Austria
\\ \email{stefan.meingast@univie.ac.at}
\and
Radcliffe Institute for Advanced Study, Harvard University, 10 Garden Street, Cambridge, MA 02138, USA
\and
Data Science @ Uni Vienna, Faculty of Earth Sciences Geography and Astronomy, University of Vienna, Austria
}

\date{Received November 11, 2018 / November 29, 2018}

\abstract{We report the discovery of two well-defined tidal tails emerging from the Hyades star cluster. The tails were detected in Gaia DR2 data by selecting cluster members in the 3D galactocentric cylindrical velocity space. The robustness of our member selection is reinforced by the fact that the sources depict an almost noiseless, coeval stellar main sequence in the observational Hertzsprung-Russel diagram. The spatial arrangement of the selected members represents a highly flattened shape with respect to the direction of movement along the clusters' orbit in the Galaxy. The size of the entire structure, within the limits of the observations, measures about \SI{200}{pc} in its largest extent, while being only about \SI{25}{pc} thick. This translates into an on-sky extent of well beyond \SI{100}{deg}. Intriguingly, a top-down view on the spatial distribution reveals a distinct S-shape, reminiscent of tidal tails that have been observed for globular clusters and also of tails that were modeled for star clusters bound to the Galactic disk. Even more remarkable, the spatial arrangement as well as the velocity dispersion of our source selection is in excellent agreement with previously published theoretical predictions for the tidal tails of the Hyades. An investigation into observed signatures of equipartition of kinetic energy, that is, mass segregation, remains unsuccessful, most likely because of the sensitivity limit for radial velocity measurements with Gaia.}

\keywords{Stars: kinematics and dynamics -- open clusters and associations: individual: Hyades}

\maketitle

\section{Introduction}
\label{sec:introduction}

The dynamical evolution of star clusters in tidal fields has seen many important contributions over the past century \citep[see, e.g.,][and references therein]{Bok34,Spitzer1940,Terlevich87,Baumgardt03}. Following theoretical and numerical approaches, clusters in galactic tidal fields become flattened perpendicular to their direction of movement over time and are also expected to develop so-called tidal tails. Unambiguous observational evidence for large-scale tidal tails has so far only been found for globular clusters \citep[e.g.,][]{Grillmair95,Dehnen04,Kaderali18}, with only some observational evidence of flattened shapes of open clusters \citep[e.g.,][]{Bergond01,Dalessandro15,Reino18}. A clear detection of extended tidal tails for clusters bound to the Galactic disk has so far been elusive.

The Hyades are the closest cluster to Earth ($d\sim$\SI{46}{pc}; \citealp{Perryman98}) and are a benchmark for stellar evolution, planet search, and cosmic distance ladder studies. The existence of tidal tails surrounding the Hyades cluster has long been predicted as the outcome of Galactic tidal forces acting on the cluster \citep[e.g.,][]{Chumak05,Chumak06,Ernst11,Roeser11}. In this letter, we first identify members of the Hyades open cluster based on kinematic data from Gaia data release 2 \citep[hereinafter referred to as Gaia DR2;][]{gaia_mission,gaia_dr2}. Based on 3D velocity data, we then present evidence for the existence of extended tidal tails associated with the cluster. Moreover, we show that the spatial distribution for stars in the tidal tails is in excellent agreement with theoretical predictions.

\section{Data}
\label{sec:data}

We used the six-dimensional position ($\alpha, \delta, \varpi $) and velocity ($\mu_\alpha$, $\mu_\delta$, rv) information from Gaia DR2 for the identification and analysis of Hyades cluster members. To obtain distances ($d$) from the measured parallaxes, we used the conversion as published by \citet{bailer-jones18}. These distance estimates are based on Bayesian inference and, generally speaking, use a position-dependent prior for the probability distribution of the distance along a particular line of sight. The mean difference in our final sample (see below) between the statistically derived distances and a straightforward $1/\varpi$ conversion amounts only \SI{0.6}{pc} or about 0.4\%, most of which originates in the global parallax zero-point of $\varpi_{\mathrm{ZP}} = \SI{-0.029}{mas}$ \citep{gaia_astrom_solution}.

\begin{figure*}[t!]
        \centering
        \resizebox{\hsize}{!}{\includegraphics[]{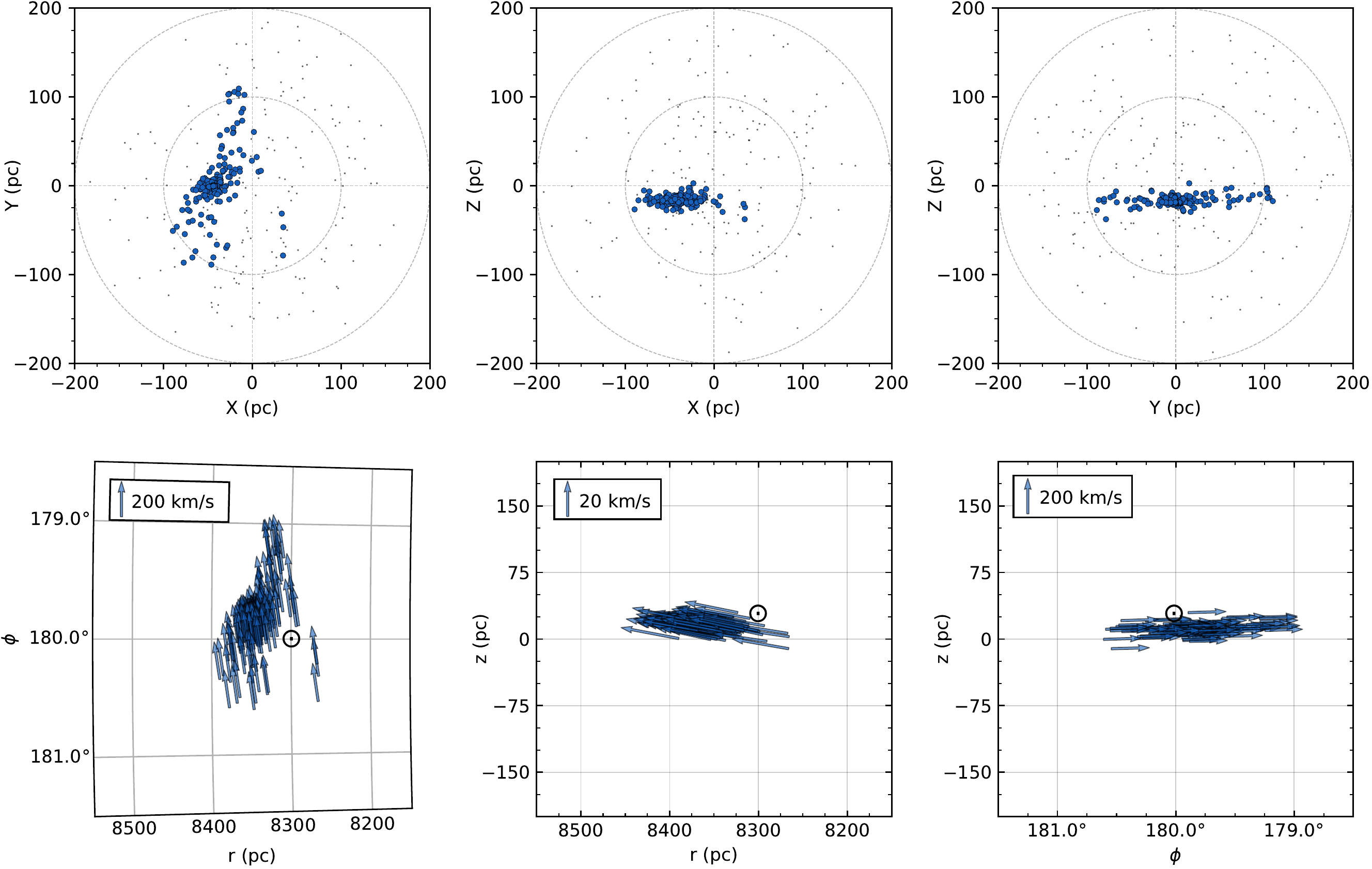}}
        \caption[]{Top row: Positions of Hyades members in Galactic Cartesian coordinates. The X-axis points toward the Galactic center, Y is positive in the direction of rotation in the Galaxy, and Z points to the Galactic poles (positive toward the north pole). The Sun is located at (0,0,0). While the fully opaque blue points refer to our final member selection, the gray dots in the background represent all sources. Bottom row: Velocity vectors for the selected Hyades members in the Galactocentric cylindrical frame. The Sun's position is indicated with the black circular symbol. The vector lengths are proportional to the velocities, as indicated in the embedded legends.}
    \label{img:xyz}
\end{figure*}

As our intention is to identify sources associated with the Hyades cluster, we limited our search to a maximum distance of \SI{200}{pc} to the Sun. Without additional constraints, limiting the database to $d \leq \SI{200}{pc}$ leaves \SI{1730451} sources in the catalog. As we require radial velocities to derive a full kinematical description, our sample is further reduced to \SI{359911} sources. In order to keep the error budget relatively small, we furthermore imposed the following quality criteria on our final sample: $\sigma_{\varpi} / \varpi < 0.1$, $\sigma_{\mu_{\alpha,\delta}} / \mu_{\alpha,\delta} < 0.1$, $\sigma_{rv} / rv < 0.1$, and max$_{\sigma\mathrm{5D}} < 0.5$. Because of our distance limit, the criteria on small parallax and proper motion errors only remove about 2\% of the initial database, while the error criterion on radial velocities accounts for the bulk of the removed sources. After applying all these limits, our final database contains \SI{298011} sources, \SI{297265} of which have valid entries for all three photometric passbands ($G$, $G_{BP}$, $G_{RP}$). These sources constitute the basis for our analysis where results and plots presented in this letter originate in this database and subsets thereof.

\section{Hyades member selection}
\label{sec:hyades_selection}

In order to identify sources associated with the Hyades cluster, we relied primarily on 3D space velocities. To optimize the extraction process, we transformed the observables into a cylindrical coordinate system with its origin at the Galactic center, that is, galactocentric cylindrical coordinates. The relevant spatial coordinates then are $(r, \phi, z),$ and they refer to the galactocentric radius, the position angle, and the height above the galactic plane, respectively. Following this convention, we used $(v_r, v_{\phi}, v_z)$ as the corresponding velocities in this coordinate frame. Naturally, this conversion requires a number of assumptions about the solar position and velocity in the Galaxy. These are listed in Appendix \ref{app:coordinates}. All coordinate transformations were carried out with well-tested functions in the \code{Astropy Python} environment. We note here that we preferred the Galactocentric cylindrical frame because the actual orbit of large extended structures is expected to show significant curvature in a Cartesian frame (making the standard Galactic Cartesian velocities $UVW$ less reliable for the identification of individual members).

Based on the computed velocities, we applied a two-step source extraction procedure to obtain sources related to the Hyades. First, we selected all sources within a radius of \SI{2.5}{\km \per \s} around the Galactocentric cylindrical velocity parameters of the Hyades: $v_r=$\;\SI{31.3}{\km \per \s}, $v_{\phi}=$\;\SI{213.0}{\km \per \s}, and $v_z=$\;\SI{6.2}{\km \per \s}. The search radius is motivated by the typical velocity dispersion of around \SI{1.5}{\km \per \s} for young stellar systems \citep[e.g.,][and references therein]{Preibisch08,riedel17}. Moreover, this characteristic can also be attributed to localized regions in tidal tails of clusters (and depending on the distance from the center), disrupting in the gravitational potential of their host galaxy \citep[e.g.,][]{Chumak10,Dehnen04}. Applying these criteria, a total of 415 sources were extracted from the Gaia DR2 database.

Second, we applied a spatial density filter by eliminating all sources with fewer than three neighbors within \SI{20}{pc}. We tried multiple combinations for these two parameters (velocity search radius and spatial density criterion) where our final choice was mainly motivated by the wish to obtain a reasonably large number of sources and at the same time very few scattered sources in the Hertzsprung-Russel diagram (HRD). The final sample consists of 238 sources out of the 415 sources that survived the first filtering stage\footnote{A catalog of selected sources will be made available via CDS}. A short discussion on the completeness of our sample is included in Appendix \ref{sec:app:completeness}.

We note here that no filtering or cleaning has been performed in any other parameter space (such as colors). Figure~\ref{img:xyz} shows all sources that were selected by this process in Cartesian Galactic coordinates centered on the Sun (top row), and also in Galactocentric cylindrical coordinates (bottom row). Figure~\ref{img:allsky} in the appendix displays the same selection in Equatorial and Galactic coordinates projected onto the sky.

\begin{figure}
        \centering
        \resizebox{0.99\hsize}{!}{\includegraphics[]{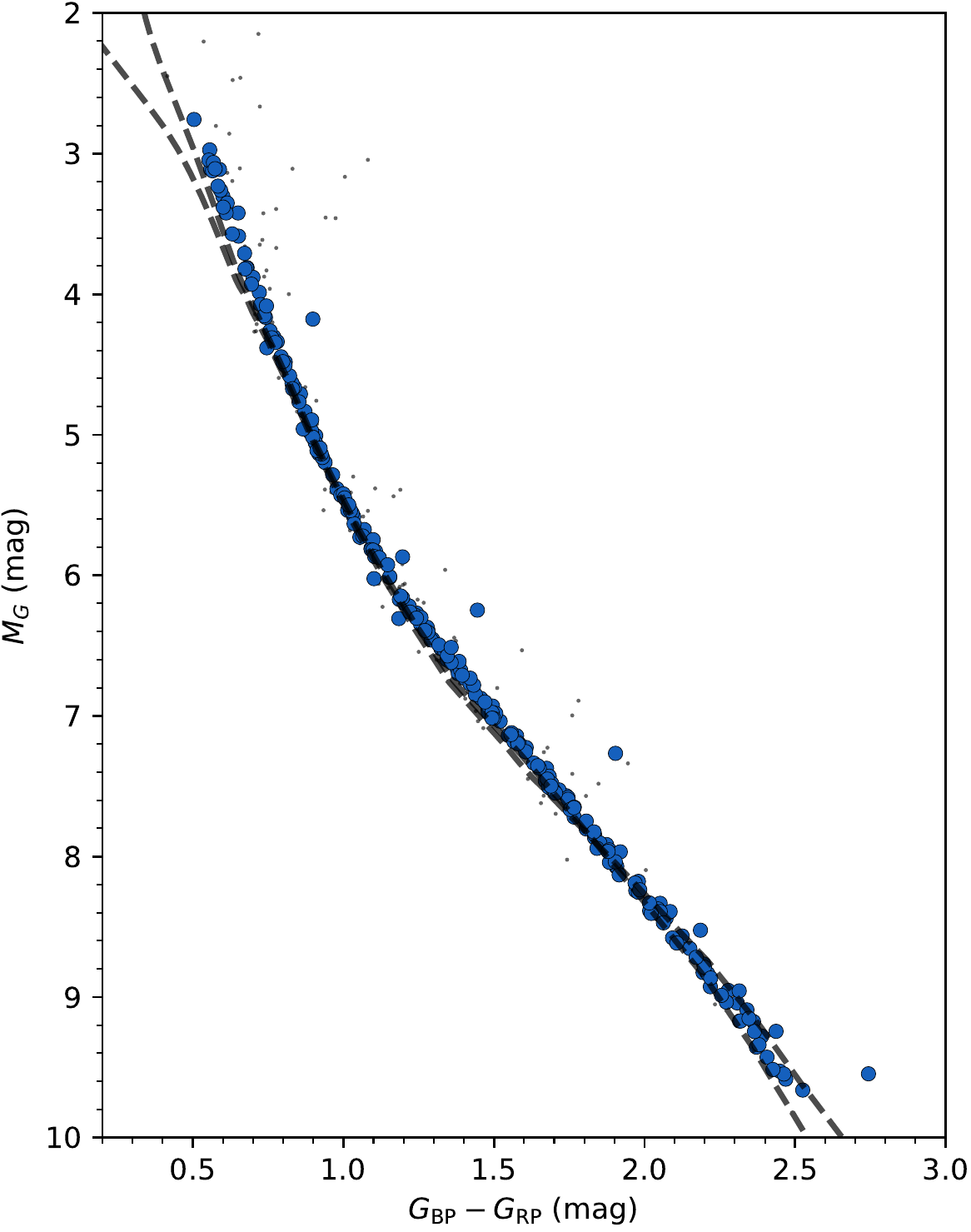}}
        \caption[]{Observational HRD for our Hyades member selection. The data and colors are the same as in Fig.\ref{img:xyz}. Two PARSEC isochrones are also shown, and both have a metallicity of Z=0.02. The top line shows an age of \SI{100}{Myr}, and the bottom line \SI{1}{Gyr}.} 
    \label{img:hrd}
\end{figure}

\begin{figure}
        \centering
        \resizebox{\hsize}{!}{\includegraphics[]{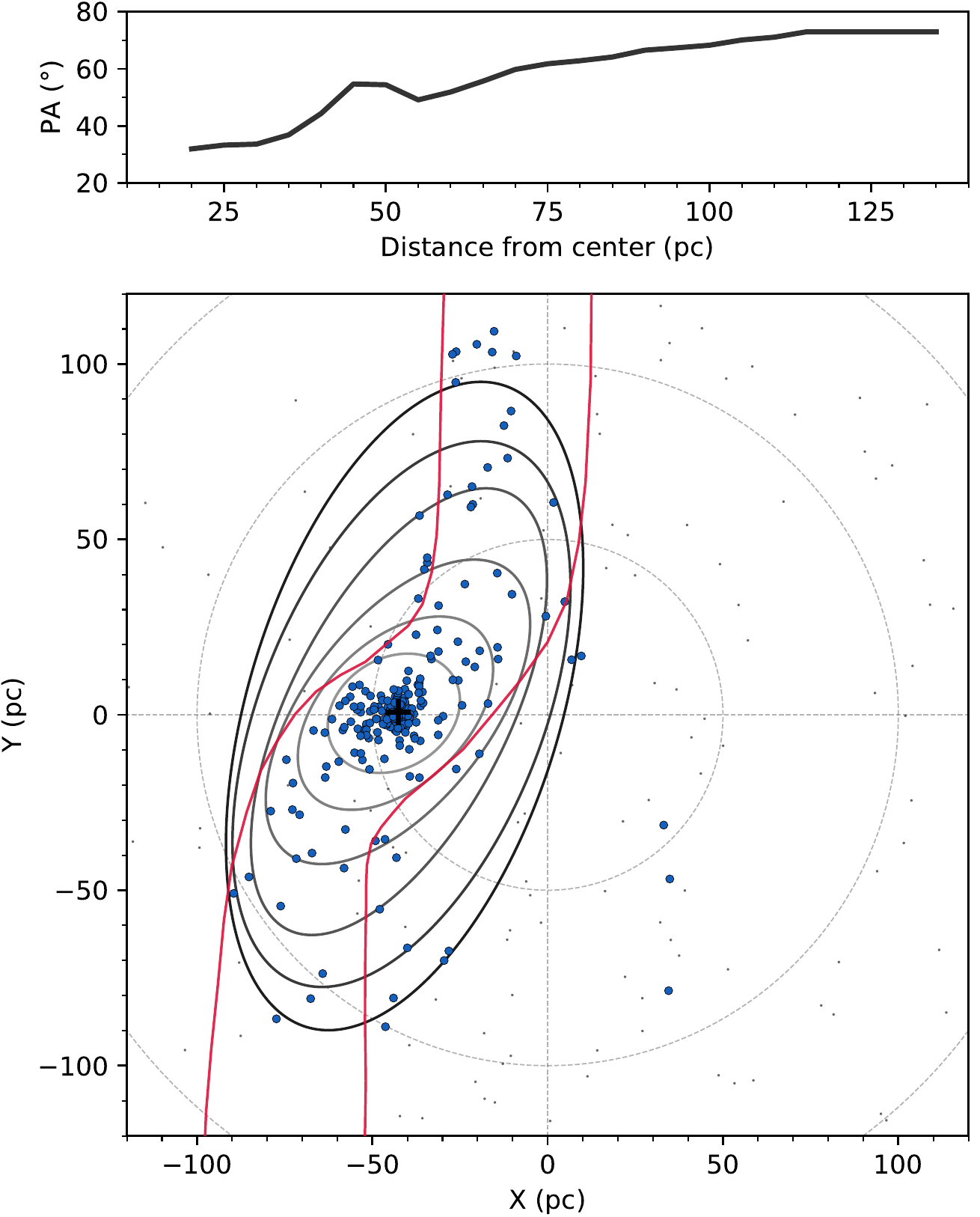}}
        \caption[]{Hyades member selection in Galactic Cartesian XY coordinates (i.e., top-down view on the Galactic plane). The data and colors are the same as in Fig.\ref{img:xyz}. We also overplot ellipses (3-$\sigma$ level computed from the covariance matrix) on the sources as determined for increasing distances to the cluster center (marked as a black cross). The top panel shows the position angle of these ellipses as a function of the distance to the center. The main panel also shows an approximate contour of the Hyades tidal tails as given in \citet{Chumak05}.}
    \label{img:hyades_tails}
\end{figure}

To check the quality of our selection, we plot an observational HRD in Fig.~\ref{img:hrd}. This figure shows that the selected sources altogether constitute an almost noiseless main sequence. Furthermore, we show two PARSEC isochrones \citep{Bressan12} for the Hyades metallicity (as in \citealp{Babusiaux18}) that fit our selection very well over a wide range of magnitudes. We also assessed whether the sequence in the HRD might be a random draw from the underlying parent distribution. To this end, we performed a two-dimensional KS test \citep{Peacock83}\footnote{We used the 2D KS-test \code{Python} implementation available at \href{https://github.com/syrte/ndtest}{https://github.com/syrte/ndtest}} to compare the 2D parameter space ($G_{\mathrm{BP}} - G_{\mathrm{BP}}$, $M_G$) for both source selections to randomly drawn subsamples of the known parent distribution. This comparison was made 100 times, and we found an average p-value of $1.6 \times 10^{-6}$ for our Hyades member selection. We also repeated this test with two randomly drawn samples for which we typically found p-values of 0.5. These findings therefore strongly suggest that the selected population is coeval and indeed constitutes a very clean selection of Hyades cluster members. 

Quite remarkably, the selected Hyades members in Fig.~\ref{img:xyz} show a clearly elongated shape. We obtained an estimate of the physical extent by computing the 3D covariance matrix of the Cartesian spatial coordinates of the selected sources (to obtain a more robust estimate, we also removed the minimum and maximum value in each dimension before computing the covariance). The resulting sizes for a 3-$\sigma$ ellipsoid in Galactic Cartesian coordinates are $\Delta X=$\;\SI{182.3}{pc}, $\Delta Y=$\;\SI{72.3}{pc}, $\Delta Z=$\;\SI{23.8}{pc}. These numbers confirm the strikingly flat configuration of the Hyades, where our selection is characterized by a rather extreme axial ratio of about 8:1. The bottom right panel in Fig.~\ref{img:xyz} reveals that the cluster is well aligned with its direction of movement in the Galaxy, not with the Galactic plane.

\section{Hyades tidal tails}
\label{sec:hyades_tails}

In this section we first describe the identification and characterization of the tidal tails. We then continue with a short discussion on mass segregation.

\subsection{Identification and characterization}

As mentioned previously, the cluster shows a highly flattened structure parallel to the Galactic plane with a rather extreme aspect ratio. However, the XY distribution in Fig.~\ref{img:xyz} shows another quite remarkable attribute. Qualitatively, the sources in the main selection seem to follow a distinct S-shape, with a central cluster core and warped extended tails. To quantitatively describe this effect, we computed the covariance matrix for subsets of the selection, based on the distance to the apparent center of the cluster. The results are shown in Fig.~\ref{img:hyades_tails}, where the main panel displays a top-down view (XY) for our Hyades selection. There, we show a series of 3-$\sigma$ ellipsoids (derived from the covariance matrix) for increasing distances from the cluster center determined as the median coordinate. The position angles (PA = $0^\circ$ would be an ellipse whose major axis points toward the Galactic center) of these ellipses are shown in the top panel as a function of distance from the cluster center. Clearly, the position angle changes with radius, and a warp of the structure becomes apparent.

To highlight the striking similarity with theoretically predicted tidal tails, we also show with red solid lines an approximate extent of the Hyades tails as given in \citet{Chumak05}. Clearly, all features point to the conclusion that our selection indeed contains many members from the Hyades tidal tails as all our findings very closely resemble the expected shape of such tails of clusters as they move through the gravitational potential of their host galaxy (specifically, see \citealp{Chumak05,Roeser11,Ernst11} for numerical results on the Hyades, but also, e.g., \citealp{Dehnen04,Chumak10} for other representations of tidal tails). The three outliers seen at X $\sim$ \SI{30}{pc} do not particularly stand out in other parameter combinations (like on the HRD) and therefore were not removed. Because they do not fit well into the general expected shape of the tails, it is still likely, however, that these are contaminating sources.

\begin{figure}
        \centering
        \resizebox{\hsize}{!}{\includegraphics[]{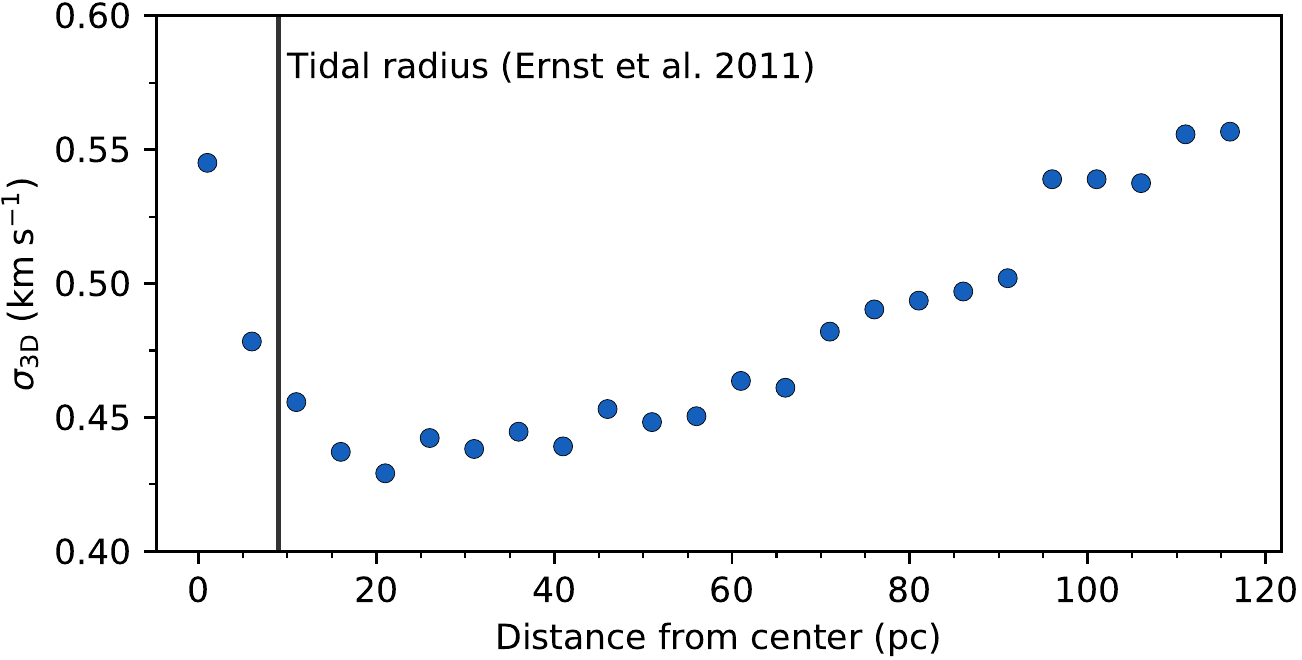}}
        \caption[]{3D velocity dispersion as a function of distance from the cluster center. The velocity dispersion was computed for all sources inside a given radius. The vertical line marks the tidal radius of \SI{9}{pc} as given in, e.g., \citet[][]{Ernst11}}
    \label{img:veldisp}
\end{figure}

Another excellent match to theoretical predictions is shown in Fig.~\ref{img:veldisp}. This plot displays the 3D velocity dispersion as a function of distance from the cluster center. Specifically, we calculated the 3D velocity dispersion for all members inside the given distance. As expected from theoretical predictions \citep{Ernst11,Roeser11}, the velocity dispersion drops when moving outward from the cluster core toward the tidal radius and again increases farther out. The absolute values of the velocity dispersion also match those quoted in \citet{Ernst11} well.

According to \citet{Ernst11}, the size of the present-day tidal tails for the Hyades should have reached a length of \SI{{\sim}800}{pc} on each side of the cluster. In contrast, in our main selection process each of the arms reaches a length of about \SI{100}{pc}. Most likely, however, our selection process introduces a bias in this measurement since the end-to-end velocity dispersion in tidal tails is expected to be much larger than our search radius. We also attempted to identify more distant sources in the tidal tail by relaxing our filter criteria. Specifically, we assessed a selection in an extended parameter space by increasing the maximum allowed distance in velocity space to \SI{3.5}{\km \per \s} from the central velocity coordinate while implementing a subsequent filtering based on distance to a given isochrone in the color-absolute magnitude space. We experimented with a wide range of parameters in the filtering setup, but in all cases, most of the added sources seemed to be randomly distributed. A better identification for such additional tail members could be achieved by matching a model of the expected tidal tail for the Hyades to our selection and then searching only for stars within these spatial limits. This is beyond the scope of this Letter, however, but would be a most interesting follow-up study. Moreover, we also expect that the sensitivity limit for radial velocity measurements in the Gaia database limits the identification process, as more remote tail members would be too faint.

\begin{figure}
        \centering
        \resizebox{\hsize}{!}{\includegraphics[]{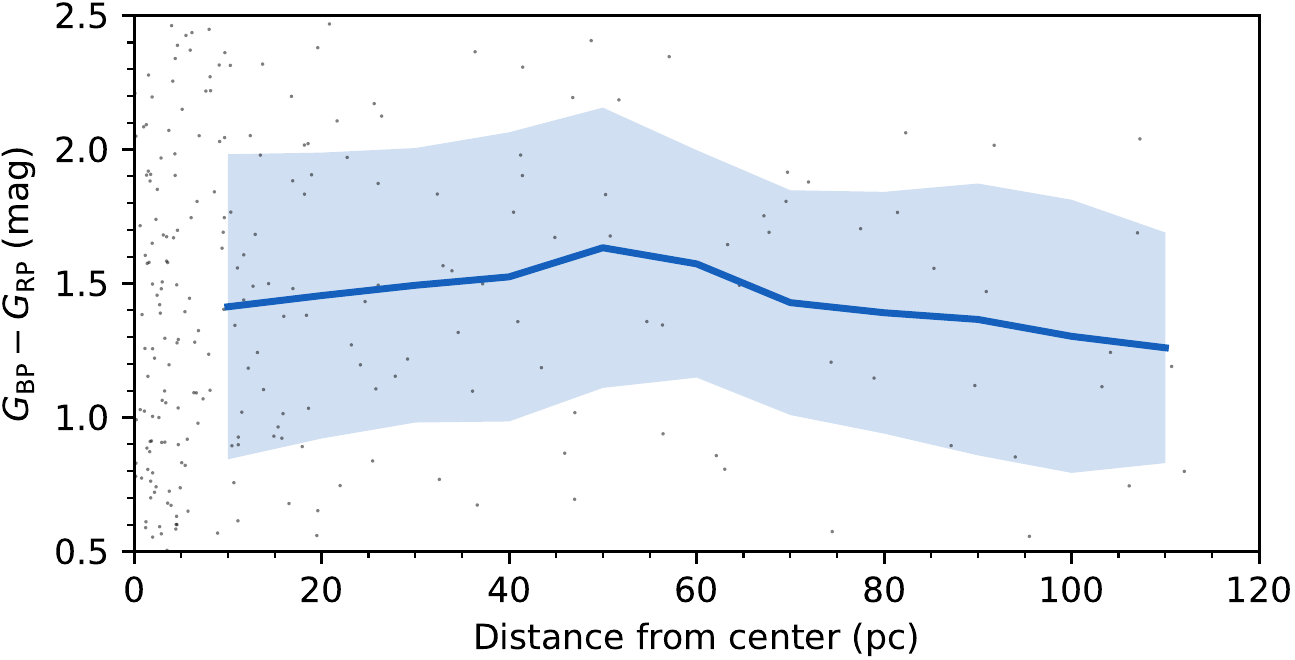}}
        \caption[]{Average $G_{\mathrm{BP}} - G_{\mathrm{RP}}$ color as a function of distance from the cluster center. The solid line shows the average and the shaded area the standard deviation in each bin. The gray dots in the background are our selected Hyades members. When we interpret the color as a proxy for stellar mass, no evidence for mass segregation can be found within the limits of our member selection and sample completeness.}
    \label{img:mass_segregation}
\end{figure}

\subsection{Mass segregation}

According to kinetic theory \citep{Binney87}, more massive stars in clusters tend to lose kinetic energy to less massive stars over time, a principle called equipartition of kinetic energy. As a consequence, massive stars are expected to move toward the center of a system, while lower-mass stars can more easily evaporate from a cluster. Thus, we would expect to preferentially find lower-mass sources in tidal tails. 

We tested this hypothesis for the Hyades tidal tails by using the Gaia color index $G_{\mathrm{BP}} - G_{\mathrm{RP}}$ as a proxy for the stellar mass (lower-mass, early-type source have redder colors than high-mass stars). Specifically, we computed the average color in consecutive annuli as a function of distance from the presumed center of the structure in Galactic Cartesian XY coordinates. The result is shown in Fig.~\ref{img:mass_segregation}, where we illustrate the average color in steps of \SI{10}{pc}, oversampled by a factor 2 (annulus size of \SI{20}{pc}). No significant change toward red colors with distance from the central position is visible. Sources in the tidal tails of the Hyades (most distant from the cluster center) even appear to be slightly bluer (more massive) on average. We attribute this characteristic to sensitivity limitations, however: Sources in the tidal tails are more distant from the Sun, which introduces a selection bias in spectral type where only intrinsically brighter stars in the tails have radial velocity measurements. We therefore find no proof of mass segregation within our current selection limits. However, as a result of completeness limits (see Appendix \ref{sec:app:completeness}), we highly encourage a reanalysis of potential mass segregation effects with later Gaia data releases.

\section{Summary}
\label{sec:summary}

We have used the recent Gaia DR2 database to identify tidal tails associated with the Hyades cluster. Initially, we applied a series of quality criteria to the catalog data, and we computed 3D space velocities in Galactocentric cylindrical coordinates for sources up to a distance of \SI{200}{pc}. Hyades members were selected by applying a \SI{2.5}{\km \per \s} velocity cut around the cluster velocity and subsequent spatial density filtering. The resulting cluster members show a distinctly flat arrangement with respect to its direction of movement in the Galaxy. The largest extent of the structure measures about \SI{200}{pc}, but it is only about \SI{25}{pc} thick. On the sky, the entire structure spans more than \SI{100}{deg}. Moreover, the sources constitute a well-defined main sequence in the observational HRD, confirming the coeval nature and robustness of our selection.

The spatial arrangement shows a remarkable characteristic: Two tidal tails extend from the cluster core and form a distinct S-shape. Moreover, we find an excellent match to previously published theoretical predictions of the tidal tail shape. The 3D velocity dispersion also matches these predictions well, with a decreasing dispersion from the cluster core toward the tidal radius and a subsequent increase in the tidal tails. An investigation on predicted mass segregation due to equipartition of kinetic energy remains unsuccessful. This is most likely due to limits in sample completeness and our member selection. Finally, an important contribution for future studies would be an identification of further tidal tails associated with other clusters. Such studies would contribute to our understanding of the evolution of young stellar systems in Galactic tidal fields and also increase our understanding of the poorly constrained Galactic gravitational potential on the plane of the Galaxy.

\begin{acknowledgements}
We wish to thank Tim de Zeeuw for helpful discussions regarding the results of this manuscript. Furthermore, we wish to thank the anonymous referee for a quick response and helpful hints.
This work has made use of data from the European Space Agency (ESA) mission {\it Gaia} (\url{https://www.cosmos.esa.int/gaia}), processed by the {\it Gaia} Data Processing and Analysis Consortium (DPAC, \url{https://www.cosmos.esa.int/web/gaia/dpac/consortium}). Funding for the DPAC has been provided by national institutions, in particular the institutions participating in the {\it Gaia} Multilateral Agreement.
This research made use of Astropy, a community-developed core Python package for Astronomy \citep{astropy}.
This research has made use of "Aladin sky atlas" developed at CDS, Strasbourg Observatory, France \citep{bonnarel00}.
We also acknowledge the various Python packages that were used in the data analysis of this work, including NumPy \citep{numpy}, SciPy \citep{scipy}, scikit-learn \citep{scikit-learn}, scikit-image \citep{scikit-image}, and Matplotlib \citep{matplotlib}.
This research has made use of the SIMBAD database operated at CDS, Strasbourg, France \citep{simbad}.
\end{acknowledgements}

\bibliography{references}

\begin{thebibliography}{39}
\expandafter\ifx\csname natexlab\endcsname\relax\def\natexlab#1{#1}\fi

\bibitem[{{Astropy Collaboration} {et~al.}(2018){Astropy Collaboration},
  {Price-Whelan}, {Sip{\H o}cz}, {G{\"u}nther}, {Lim}, {Crawford}, {Conseil},
  {Shupe}, {Craig}, {Dencheva}, {Ginsburg}, {VanderPlas}, {Bradley},
  {P{\'e}rez-Su{\'a}rez}, {de Val-Borro}, {Aldcroft}, {Cruz}, {Robitaille},
  {Tollerud}, {Ardelean}, {Babej}, {Bach}, {Bachetti}, {Bakanov}, {Bamford},
  {Barentsen}, {Barmby}, {Baumbach}, {Berry}, {Biscani}, {Boquien}, {Bostroem},
  {Bouma}, {Brammer}, {Bray}, {Breytenbach}, {Buddelmeijer}, {Burke},
  {Calderone}, {Cano Rodr{\'{\i}}guez}, {Cara}, {Cardoso}, {Cheedella},
  {Copin}, {Corrales}, {Crichton}, {D'Avella}, {Deil}, {Depagne}, {Dietrich},
  {Donath}, {Droettboom}, {Earl}, {Erben}, {Fabbro}, {Ferreira}, {Finethy},
  {Fox}, {Garrison}, {Gibbons}, {Goldstein}, {Gommers}, {Greco}, {Greenfield},
  {Groener}, {Grollier}, {Hagen}, {Hirst}, {Homeier}, {Horton}, {Hosseinzadeh},
  {Hu}, {Hunkeler}, {Ivezi{\'c}}, {Jain}, {Jenness}, {Kanarek}, {Kendrew},
  {Kern}, {Kerzendorf}, {Khvalko}, {King}, {Kirkby}, {Kulkarni}, {Kumar},
  {Lee}, {Lenz}, {Littlefair}, {Ma}, {Macleod}, {Mastropietro}, {McCully},
  {Montagnac}, {Morris}, {Mueller}, {Mumford}, {Muna}, {Murphy}, {Nelson},
  {Nguyen}, {Ninan}, {N{\"o}the}, {Ogaz}, {Oh}, {Parejko}, {Parley}, {Pascual},
  {Patil}, {Patil}, {Plunkett}, {Prochaska}, {Rastogi}, {Reddy Janga},
  {Sabater}, {Sakurikar}, {Seifert}, {Sherbert}, {Sherwood-Taylor}, {Shih},
  {Sick}, {Silbiger}, {Singanamalla}, {Singer}, {Sladen}, {Sooley},
  {Sornarajah}, {Streicher}, {Teuben}, {Thomas}, {Tremblay}, {Turner},
  {Terr{\'o}n}, {van Kerkwijk}, {de la Vega}, {Watkins}, {Weaver}, {Whitmore},
  {Woillez}, {Zabalza}, \& {Astropy Contributors}}]{astropy}
{Astropy Collaboration}, {Price-Whelan}, A.~M., {Sip{\H o}cz}, B.~M., {et~al.}
  2018, \aj, 156, 123

\bibitem[{{Babusiaux} {et~al.}(2018){Babusiaux}, {van Leeuwen}, {Barstow},
  {Jordi}, {Vallenari}, {Bossini}, {Bressan}, {Cantat-Gaudin}, {van Leeuwen},
  {Brown}, {Prusti}, {de Bruijne}, {Bailer-Jones}, {Biermann}, {Evans}, {Eyer},
  {Jansen}, {Klioner}, {Lammers}, {Lindegren}, {Luri}, {Mignard}, {Panem},
  {Pourbaix}, {Randich}, {Sartoretti}, {Siddiqui}, {Soubiran}, {Walton},
  {Arenou}, {Bastian}, {Cropper}, {Drimmel}, {Katz}, {Lattanzi}, {Bakker},
  {Cacciari}, {Casta{\~n}eda}, {Chaoul}, {Cheek}, {De Angeli}, {Fabricius},
  {Guerra}, {Holl}, {Masana}, {Messineo}, {Mowlavi}, {Nienartowicz}, {Panuzzo},
  {Portell}, {Riello}, {Seabroke}, {Tanga}, {Th{\'e}venin}, {Gracia-Abril},
  {Comoretto}, {Garcia-Reinaldos}, {Teyssier}, {Altmann}, {Andrae}, {Audard},
  {Bellas- Velidis}, {Benson}, {Berthier}, {Blomme}, {Burgess}, {Busso},
  {Carry}, {Cellino}, {Clementini}, {Clotet}, {Creevey}, {Davidson}, {De
  Ridder}, {Delchambre}, {Dell'Oro}, {Ducourant},
  {Fern{\'a}ndez-Hern{\'a}ndez}, {Fouesneau}, {Fr{\'e}mat}, {Galluccio},
  {Garc{\'\i}a-Torres}, {Gonz{\'a}lez-N{\'u}{\~n}ez}, {Gonz{\'a}lez- Vidal},
  {Gosset}, {Guy}, {Halbwachs}, {Hambly}, {Harrison}, {Hern{\'a}ndez},
  {Hestroffer}, {Hodgkin}, {Hutton}, {Jasniewicz}, {Jean-Antoine- Piccolo},
  {Jordan}, {Korn}, {Krone- Martins}, {Lanzafame}, {Lebzelter}, {L{\"o}ffler},
  {Manteiga}, {Marrese}, {Mart{\'\i}n-Fleitas}, {Moitinho}, {Mora}, {Muinonen},
  {Osinde}, {Pancino}, {Pauwels}, {Petit}, {Recio-Blanco}, {Richards},
  {Rimoldini}, {Robin}, {Sarro}, {Siopis}, {Smith}, {Sozzetti}, {S{\"u}veges},
  {Torra}, {van Reeven}, {Abbas}, {Abreu Aramburu}, {Accart}, {Aerts},
  {Altavilla}, {{\'A}lvarez}, {Alvarez}, {Alves}, {Anderson}, {Andrei},
  {Anglada Varela}, {Antiche}, {Antoja}, {Arcay}, {Astraatmadja}, {Bach},
  {Baker}, {Balaguer-N{\'u}{\~n}ez}, {Balm}, {Barache}, {Barata}, {Barbato},
  {Barblan}, {Barklem}, {Barrado}, {Barros}, {Bartholom{\'e} Mu{\~n}oz},
  {Bassilana}, {Becciani}, {Bellazzini}, {Berihuete}, {Bertone}, {Bianchi},
  {Bienaym{\'e}}, {Blanco-Cuaresma}, {Boch}, {Boeche}, {Bombrun}, {Borrachero},
  {Bouquillon}, {Bourda}, {Bragaglia}, {Bramante}, {Breddels}, {Brouillet},
  {Br{\"u}semeister}, {Brugaletta}, {Bucciarelli}, {Burlacu}, {Busonero},
  {Butkevich}, {Buzzi}, {Caffau}, {Cancelliere}, {Cannizzaro}, {Carballo},
  {Carlucci}, {Carrasco}, {Casamiquela}, {Castellani}, {Castro-Ginard},
  {Charlot}, {Chemin}, {Chiavassa}, {Cocozza}, {Costigan}, {Cowell}, {Crifo},
  {Crosta}, {Crowley}, {Cuypers}, {Dafonte}, {Damerdji}, {Dapergolas}, {David},
  {David}, {de Laverny}, {De Luise}, {De March}, {de Martino}, {de Souza}, {de
  Torres}, {Debosscher}, {del Pozo}, {Delbo}, {Delgado}, {Delgado}, {Diakite},
  {Diener}, {Distefano}, {Dolding}, {Drazinos}, {Dur{\'a}n}, {Edvardsson},
  {Enke}, {Eriksson}, {Esquej}, {Eynard Bontemps}, {Fabre}, {Fabrizio},
  {Faigler}, {Falc{\~a}o}, {Farr{\`a}s Casas}, {Federici}, {Fedorets},
  {Fernique}, {Figueras}, {Filippi}, {Findeisen}, {Fonti}, {Fraile}, {Fraser},
  {Fr{\'e}zouls}, {Gai}, {Galleti}, {Garabato}, {Garc{\'\i}a-Sedano},
  {Garofalo}, {Garralda}, {Gavel}, {Gavras}, {Gerssen}, {Geyer}, {Giacobbe},
  {Gilmore}, {Girona}, {Giuffrida}, {Glass}, {Gomes}, {Granvik}, {Gueguen},
  {Guerrier}, {Guiraud}, {Guti{\'e}}, {Haigron}, {Hatzidimitriou}, {Hauser},
  {Haywood}, {Heiter}, {Helmi}, {Heu}, {Hilger}, {Hobbs}, {Hofmann}, {Holland},
  {Huckle}, {Hypki}, {Icardi}, {Jan{\ss}en}, {Jevardat de Fombelle}, {Jonker},
  {Juh{\'a}sz}, {Julbe}, {Karampelas}, {Kewley}, {Klar}, {Kochoska}, {Kohley},
  {Kolenberg}, {Kontizas}, {Kontizas}, {Koposov}, {Kordopatis},
  {Kostrzewa-Rutkowska}, {Koubsky}, {Lambert}, {Lanza}, {Lasne}, {Lavigne}, {Le
  Fustec}, {Le Poncin-Lafitte}, {Lebreton}, {Leccia}, {Leclerc},
  {Lecoeur-Taibi}, {Lenhardt}, {Leroux}, {Liao}, {Licata}, {Lindstr{\o}m},
  {Lister}, {Livanou}, {Lobel}, {L{\'o}pez}, {Managau}, {Mann}, {Mantelet},
  {Marchal}, {Marchant}, {Marconi}, {Marinoni}, {Marschalk{\'o}}, {Marshall},
  {Martino}, {Marton}, {Mary}, {Massari}, {Matijevi{\v{c}}}, {Mazeh},
  {McMillan}, {Messina}, {Michalik}, {Millar}, {Molina}, {Molinaro},
  {Moln{\'a}r}, {Montegriffo}, {Mor}, {Morbidelli}, {Morel}, {Morris},
  {Mulone}, {Muraveva}, {Musella}, {Nelemans}, {Nicastro}, {Noval},
  {O'Mullane}, {Ord{\'e}novic}, {Ord{\'o}{\~n}ez-Blanco}, {Osborne}, {Pagani},
  {Pagano}, {Pailler}, {Palacin}, {Palaversa}, {Panahi}, {Pawlak},
  {Piersimoni}, {Pineau}, {Plachy}, {Plum}, {Poggio}, {Poujoulet},
  {Pr{\v{s}}a}, {Pulone}, {Racero}, {Ragaini}, {Rambaux}, {Ramos-Lerate},
  {Regibo}, {Reyl{\'e}}, {Riclet}, {Ripepi}, {Riva}, {Rivard}, {Rixon},
  {Roegiers}, {Roelens}, {Romero-G{\'o}mez}, {Rowell}, {Royer}, {Ruiz-Dern},
  {Sadowski}, {Sagrist{\`a} Sell{\'e}s}, {Sahlmann}, {Salgado}, {Salguero},
  {Sanna}, {Santana- Ros}, {Sarasso}, {Savietto}, {Schultheis}, {Sciacca},
  {Segol}, {Segovia}, {S{\'e}gransan}, {Shih}, {Siltala}, {Silva}, {Smart},
  {Smith}, {Solano}, {Solitro}, {Sordo}, {Soria Nieto}, {Souchay}, {Spagna},
  {Spoto}, {Stampa}, {Steele}, {Steidelm{\"u}ller}, {Stephenson}, {Stoev},
  {Suess}, {Surdej}, {Szabados}, {Szegedi-Elek}, {Tapiador}, {Taris}, {Tauran},
  {Taylor}, {Teixeira}, {Terrett}, {Teyssandier}, {Thuillot}, {Titarenko},
  {Torra Clotet}, {Turon}, {Ulla}, {Utrilla}, {Uzzi}, {Vaillant}, {Valentini},
  {Valette}, {van Elteren}, {Van Hemelryck}, {Vaschetto}, {Vecchiato},
  {Veljanoski}, {Viala}, {Vicente}, {Vogt}, {von Essen}, {Voss}, {Votruba},
  {Voutsinas}, {Walmsley}, {Weiler}, {Wertz}, {Wevers}, {Wyrzykowski},
  {Yoldas}, {{\v{Z}}erjal}, {Ziaeepour}, {Zorec}, {Zschocke}, {Zucker},
  {Zurbach}, \& {Zwitter}}]{Babusiaux18}
{Babusiaux}, C., {van Leeuwen}, F., {Barstow}, M.~A., {et~al.} 2018, \aap, 616,
  A10

\bibitem[{{Bailer-Jones} {et~al.}(2018){Bailer-Jones}, {Rybizki}, {Fouesneau},
  {Mantelet}, \& {Andrae}}]{bailer-jones18}
{Bailer-Jones}, C.~A.~L., {Rybizki}, J., {Fouesneau}, M., {Mantelet}, G., \&
  {Andrae}, R. 2018, ArXiv e-prints, arXiv:1804.10121

\bibitem[{{Baumgardt} \& {Makino}(2003)}]{Baumgardt03}
{Baumgardt}, H. \& {Makino}, J. 2003, \mnras, 340, 227

\bibitem[{{Bergond} {et~al.}(2001){Bergond}, {Leon}, \& {Guibert}}]{Bergond01}
{Bergond}, G., {Leon}, S., \& {Guibert}, J. 2001, \aap, 377, 462

\bibitem[{{Binney} \& {Tremaine}(1987)}]{Binney87}
{Binney}, J. \& {Tremaine}, S. 1987, {Galactic Dynamics} (Princeton University
  Press)

\bibitem[{{Bok}(1934)}]{Bok34}
{Bok}, B.~J. 1934, Harvard College Observatory Circular, 384, 1

\bibitem[{{Bonnarel} {et~al.}(2000){Bonnarel}, {Fernique}, {Bienaym{\'e}},
  {Egret}, {Genova}, {Louys}, {Ochsenbein}, {Wenger}, \&
  {Bartlett}}]{bonnarel00}
{Bonnarel}, F., {Fernique}, P., {Bienaym{\'e}}, O., {et~al.} 2000, \aaps, 143,
  33

\bibitem[{{Bressan} {et~al.}(2012){Bressan}, {Marigo}, {Girardi}, {Salasnich},
  {Dal Cero}, {Rubele}, \& {Nanni}}]{Bressan12}
{Bressan}, A., {Marigo}, P., {Girardi}, L., {et~al.} 2012, \mnras, 427, 127

\bibitem[{{Brown} {et~al.}(2018){Brown}, {Vallenari}, {Prusti}, {de Bruijne},
  {Babusiaux}, {Bailer-Jones}, {Biermann}, {Evans}, {Eyer}, {Jansen}, {Jordi},
  {Klioner}, {Lammers}, {Lindegren}, {Luri}, {Mignard}, {Panem}, {Pourbaix},
  {Randich}, {Sartoretti}, {Siddiqui}, {Soubiran}, {van Leeuwen}, {Walton},
  {Arenou}, {Bastian}, {Cropper}, {Drimmel}, {Katz}, {Lattanzi}, {Bakker},
  {Cacciari}, {Casta{\~n}eda}, {Chaoul}, {Cheek}, {De Angeli}, {Fabricius},
  {Guerra}, {Holl}, {Masana}, {Messineo}, {Mowlavi}, {Nienartowicz}, {Panuzzo},
  {Portell}, {Riello}, {Seabroke}, {Tanga}, {Th{\'e}venin}, {Gracia-Abril},
  {Comoretto}, {Garcia-Reinaldos}, {Teyssier}, {Altmann}, {Andrae}, {Audard},
  {Bellas-Velidis}, {Benson}, {Berthier}, {Blomme}, {Burgess}, {Busso},
  {Carry}, {Cellino}, {Clementini}, {Clotet}, {Creevey}, {Davidson}, {De
  Ridder}, {Delchambre}, {Dell'Oro}, {Ducourant},
  {Fern{\'a}ndez-Hern{\'a}ndez}, {Fouesneau}, {Fr{\'e}mat}, {Galluccio},
  {Garc{\'\i}a-Torres}, {Gonz{\'a}lez-N{\'u}{\~n}ez}, {Gonz{\'a}lez- Vidal},
  {Gosset}, {Guy}, {Halbwachs}, {Hambly}, {Harrison}, {Hern{\'a}ndez},
  {Hestroffer}, {Hodgkin}, {Hutton}, {Jasniewicz}, {Jean-Antoine- Piccolo},
  {Jordan}, {Korn}, {Krone- Martins}, {Lanzafame}, {Lebzelter}, {L{\"o}ffler},
  {Manteiga}, {Marrese}, {Mart{\'\i}n-Fleitas}, {Moitinho}, {Mora}, {Muinonen},
  {Osinde}, {Pancino}, {Pauwels}, {Petit}, {Recio-Blanco}, {Richards},
  {Rimoldini}, {Robin}, {Sarro}, {Siopis}, {Smith}, {Sozzetti}, {S{\"u}veges},
  {Torra}, {van Reeven}, {Abbas}, {Abreu Aramburu}, {Accart}, {Aerts},
  {Altavilla}, {{\'A}lvarez}, {Alvarez}, {Alves}, {Anderson}, {Andrei},
  {Anglada Varela}, {Antiche}, {Antoja}, {Arcay}, {Astraatmadja}, {Bach},
  {Baker}, {Balaguer-N{\'u}{\~n}ez}, {Balm}, {Barache}, {Barata}, {Barbato},
  {Barblan}, {Barklem}, {Barrado}, {Barros}, {Barstow}, {Bartholom{\'e}
  Mu{\~n}oz}, {Bassilana}, {Becciani}, {Bellazzini}, {Berihuete}, {Bertone},
  {Bianchi}, {Bienaym{\'e}}, {Blanco-Cuaresma}, {Boch}, {Boeche}, {Bombrun},
  {Borrachero}, {Bossini}, {Bouquillon}, {Bourda}, {Bragaglia}, {Bramante},
  {Breddels}, {Bressan}, {Brouillet}, {Br{\"u}semeister}, {Brugaletta},
  {Bucciarelli}, {Burlacu}, {Busonero}, {Butkevich}, {Buzzi}, {Caffau},
  {Cancelliere}, {Cannizzaro}, {Cantat-Gaudin}, {Carballo}, {Carlucci},
  {Carrasco}, {Casamiquela}, {Castellani}, {Castro-Ginard}, {Charlot},
  {Chemin}, {Chiavassa}, {Cocozza}, {Costigan}, {Cowell}, {Crifo}, {Crosta},
  {Crowley}, {Cuypers}, {Dafonte}, {Damerdji}, {Dapergolas}, {David}, {David},
  {de Laverny}, {De Luise}, {De March}, {de Martino}, {de Souza}, {de Torres},
  {Debosscher}, {del Pozo}, {Delbo}, {Delgado}, {Delgado}, {Di Matteo},
  {Diakite}, {Diener}, {Distefano}, {Dolding}, {Drazinos}, {Dur{\'a}n},
  {Edvardsson}, {Enke}, {Eriksson}, {Esquej}, {Eynard Bontemps}, {Fabre},
  {Fabrizio}, {Faigler}, {Falc{\~a}o}, {Farr{\`a}s Casas}, {Federici},
  {Fedorets}, {Fernique}, {Figueras}, {Filippi}, {Findeisen}, {Fonti},
  {Fraile}, {Fraser}, {Fr{\'e}zouls}, {Gai}, {Galleti}, {Garabato},
  {Garc{\'\i}a-Sedano}, {Garofalo}, {Garralda}, {Gavel}, {Gavras}, {Gerssen},
  {Geyer}, {Giacobbe}, {Gilmore}, {Girona}, {Giuffrida}, {Glass}, {Gomes},
  {Granvik}, {Gueguen}, {Guerrier}, {Guiraud}, {Guti{\'e}rrez-S{\'a}nchez},
  {Haigron}, {Hatzidimitriou}, {Hauser}, {Haywood}, {Heiter}, {Helmi}, {Heu},
  {Hilger}, {Hobbs}, {Hofmann}, {Holland}, {Huckle}, {Hypki}, {Icardi},
  {Jan{\ss}en}, {Jevardat de Fombelle}, {Jonker}, {Juh{\'a}sz}, {Julbe},
  {Karampelas}, {Kewley}, {Klar}, {Kochoska}, {Kohley}, {Kolenberg},
  {Kontizas}, {Kontizas}, {Koposov}, {Kordopatis}, {Kostrzewa-Rutkowska},
  {Koubsky}, {Lambert}, {Lanza}, {Lasne}, {Lavigne}, {Le Fustec}, {Le
  Poncin-Lafitte}, {Lebreton}, {Leccia}, {Leclerc}, {Lecoeur-Taibi},
  {Lenhardt}, {Leroux}, {Liao}, {Licata}, {Lindstr{\o}m}, {Lister}, {Livanou},
  {Lobel}, {L{\'o}pez}, {Managau}, {Mann}, {Mantelet}, {Marchal}, {Marchant},
  {Marconi}, {Marinoni}, {Marschalk{\'o}}, {Marshall}, {Martino}, {Marton},
  {Mary}, {Massari}, {Matijevi{\v{c}}}, {Mazeh}, {McMillan}, {Messina},
  {Michalik}, {Millar}, {Molina}, {Molinaro}, {Moln{\'a}r}, {Montegriffo},
  {Mor}, {Morbidelli}, {Morel}, {Morris}, {Mulone}, {Muraveva}, {Musella},
  {Nelemans}, {Nicastro}, {Noval}, {O'Mullane}, {Ord{\'e}novic},
  {Ord{\'o}{\~n}ez-Blanco}, {Osborne}, {Pagani}, {Pagano}, {Pailler},
  {Palacin}, {Palaversa}, {Panahi}, {Pawlak}, {Piersimoni}, {Pineau}, {Plachy},
  {Plum}, {Poggio}, {Poujoulet}, {Pr{\v{s}}a}, {Pulone}, {Racero}, {Ragaini},
  {Rambaux}, {Ramos-Lerate}, {Regibo}, {Reyl{\'e}}, {Riclet}, {Ripepi}, {Riva},
  {Rivard}, {Rixon}, {Roegiers}, {Roelens}, {Romero-G{\'o}mez}, {Rowell},
  {Royer}, {Ruiz-Dern}, {Sadowski}, {Sagrist{\`a} Sell{\'e}s}, {Sahlmann},
  {Salgado}, {Salguero}, {Sanna}, {Santana- Ros}, {Sarasso}, {Savietto},
  {Schultheis}, {Sciacca}, {Segol}, {Segovia}, {S{\'e}gransan}, {Shih},
  {Siltala}, {Silva}, {Smart}, {Smith}, {Solano}, {Solitro}, {Sordo}, {Soria
  Nieto}, {Souchay}, {Spagna}, {Spoto}, {Stampa}, {Steele},
  {Steidelm{\"u}ller}, {Stephenson}, {Stoev}, {Suess}, {Surdej}, {Szabados},
  {Szegedi-Elek}, {Tapiador}, {Taris}, {Tauran}, {Taylor}, {Teixeira},
  {Terrett}, {Teyssandier}, {Thuillot}, {Titarenko}, {Torra Clotet}, {Turon},
  {Ulla}, {Utrilla}, {Uzzi}, {Vaillant}, {Valentini}, {Valette}, {van Elteren},
  {Van Hemelryck}, {van Leeuwen}, {Vaschetto}, {Vecchiato}, {Veljanoski},
  {Viala}, {Vicente}, {Vogt}, {von Essen}, {Voss}, {Votruba}, {Voutsinas},
  {Walmsley}, {Weiler}, {Wertz}, {Wevers}, {Wyrzykowski}, {Yoldas},
  {{\v{Z}}erjal}, {Ziaeepour}, {Zorec}, {Zschocke}, {Zucker}, {Zurbach}, \&
  {Zwitter}}]{gaia_dr2}
{Brown}, A.~G.~A., {Vallenari}, A., {Prusti}, T., {et~al.} 2018, \aap, 616, A1

\bibitem[{{Chen} {et~al.}(2001){Chen}, {Stoughton}, {Smith}, {Uomoto}, {Pier},
  {Yanny}, {Ivezi{\'c}}, {York}, {Anderson}, {Annis}, {Brinkmann}, {Csabai},
  {Fukugita}, {Hindsley}, {Lupton}, {Munn}, \& {SDSS Collaboration}}]{Chen01}
{Chen}, B., {Stoughton}, C., {Smith}, J.~A., {et~al.} 2001, \apj, 553, 184

\bibitem[{{Chumak} {et~al.}(2010){Chumak}, {Platais}, {McLaughlin},
  {Rastorguev}, \& {Chumak}}]{Chumak10}
{Chumak}, Y.~O., {Platais}, I., {McLaughlin}, D.~E., {Rastorguev}, A.~S., \&
  {Chumak}, O.~V. 2010, \mnras, 402, 1841

\bibitem[{{Chumak} \& {Rastorguev}(2006)}]{Chumak06}
{Chumak}, Y.~O. \& {Rastorguev}, A.~S. 2006, Astronomy Letters, 32, 446

\bibitem[{{Chumak} {et~al.}(2005){Chumak}, {Rastorguev}, \&
  {Aarseth}}]{Chumak05}
{Chumak}, Y.~O., {Rastorguev}, A.~S., \& {Aarseth}, S.~J. 2005, Astronomy
  Letters, 31, 308

\bibitem[{{Dalessandro} {et~al.}(2015){Dalessandro}, {Miocchi}, {Carraro},
  {J{\'\i}lkov{\'a}}, \& {Moitinho}}]{Dalessandro15}
{Dalessandro}, E., {Miocchi}, P., {Carraro}, G., {J{\'\i}lkov{\'a}}, L., \&
  {Moitinho}, A. 2015, \mnras, 449, 1811

\bibitem[{{Dehnen} {et~al.}(2004){Dehnen}, {Odenkirchen}, {Grebel}, \&
  {Rix}}]{Dehnen04}
{Dehnen}, W., {Odenkirchen}, M., {Grebel}, E.~K., \& {Rix}, H.-W. 2004, \aj,
  127, 2753

\bibitem[{{Ernst} {et~al.}(2011){Ernst}, {Just}, {Berczik}, \&
  {Olczak}}]{Ernst11}
{Ernst}, A., {Just}, A., {Berczik}, P., \& {Olczak}, C. 2011, \aap, 536, A64

\bibitem[{{Gillessen} {et~al.}(2009){Gillessen}, {Eisenhauer}, {Trippe},
  {Alexander}, {Genzel}, {Martins}, \& {Ott}}]{Gillessen09}
{Gillessen}, S., {Eisenhauer}, F., {Trippe}, S., {et~al.} 2009, \apj, 692, 1075

\bibitem[{{Grillmair} {et~al.}(1995){Grillmair}, {Freeman}, {Irwin}, \&
  {Quinn}}]{Grillmair95}
{Grillmair}, C.~J., {Freeman}, K.~C., {Irwin}, M., \& {Quinn}, P.~J. 1995, \aj,
  109, 2553

\bibitem[{Hunter(2007)}]{matplotlib}
Hunter, J.~D. 2007, Computing In Science \& Engineering, 9, 90

\bibitem[{Jones {et~al.}(2001)Jones, Oliphant, Peterson, {et~al.}}]{scipy}
Jones, E., Oliphant, T., Peterson, P., {et~al.} 2001, {SciPy}: Open source
  scientific tools for {Python}

\bibitem[{{Kaderali} {et~al.}(2018){Kaderali}, {Hunt}, {Webb}, {Price-Jones},
  \& {Carlberg}}]{Kaderali18}
{Kaderali}, S., {Hunt}, J. A.~S., {Webb}, J.~J., {Price-Jones}, N., \&
  {Carlberg}, R. 2018, ArXiv e-prints, arXiv:1809.04108

\bibitem[{{Kerr} \& {Lynden-Bell}(1986)}]{Kerr86}
{Kerr}, F.~J. \& {Lynden-Bell}, D. 1986, \mnras, 221, 1023

\bibitem[{{Lindegren} {et~al.}(2018){Lindegren}, {Hernandez}, {Bombrun},
  {Klioner}, {Bastian}, {Ramos-Lerate}, {de Torres}, {Steidelmuller},
  {Stephenson}, {Hobbs}, {Lammers}, {Biermann}, {Geyer}, {Hilger}, {Michalik},
  {Stampa}, {McMillan}, {Castaneda}, {Clotet}, {Comoretto}, {Davidson},
  {Fabricius}, {Gracia}, {Hambly}, {Hutton}, {Mora}, {Portell}, {van Leeuwen},
  {Abbas}, {Abreu}, {Altmann}, {Andrei}, {Anglada}, {Balaguer- Nunez},
  {Barache}, {Becciani}, {Bertone}, {Bianchi}, {Bouquillon}, {Bourda},
  {Brusemeister}, {Bucciarelli}, {Busonero}, {Buzzi}, {Cancelliere},
  {Carlucci}, {Charlot}, {Cheek}, {Crosta}, {Crowley}, {de Bruijne}, {de
  Felice}, {Drimmel}, {Esquej}, {Fienga}, {Fraile}, {Gai}, {Garralda},
  {Gonzalez-Vidal}, {Guerra}, {Hauser}, {Hofmann}, {Holl}, {Jordan},
  {Lattanzi}, {Lenhardt}, {Liao}, {Licata}, {Lister}, {Loffler}, {Marchant},
  {Martin-Fleitas}, {Messineo}, {Mignard}, {Morbidelli}, {Poggio}, {Riva},
  {Rowell}, {Salguero}, {Sarasso}, {Sciacca}, {Siddiqui}, {Smart}, {Spagna},
  {Steele}, {Taris}, {Torra}, {van Elteren}, {van Reeven}, \&
  {Vecchiato}}]{gaia_astrom_solution}
{Lindegren}, L., {Hernandez}, J., {Bombrun}, A., {et~al.} 2018, ArXiv e-prints,
  arXiv:1804.09366

\bibitem[{{Peacock}(1983)}]{Peacock83}
{Peacock}, J.~A. 1983, \mnras, 202, 615

\bibitem[{Pedregosa {et~al.}(2011)Pedregosa, Varoquaux, Gramfort, Michel,
  Thirion, Grisel, Blondel, Prettenhofer, Weiss, Dubourg, Vanderplas, Passos,
  Cournapeau, Brucher, Perrot, \& Duchesnay}]{scikit-learn}
Pedregosa, F., Varoquaux, G., Gramfort, A., {et~al.} 2011, Journal of Machine
  Learning Research, 12, 2825

\bibitem[{{Perryman} {et~al.}(1998){Perryman}, {Brown}, {Lebreton}, {Gomez},
  {Turon}, {Cayrel de Strobel}, {Mermilliod}, {Robichon}, {Kovalevsky}, \&
  {Crifo}}]{Perryman98}
{Perryman}, M.~A.~C., {Brown}, A.~G.~A., {Lebreton}, Y., {et~al.} 1998, \aap,
  331, 81

\bibitem[{{Preibisch} \& {Mamajek}(2008)}]{Preibisch08}
{Preibisch}, T. \& {Mamajek}, E. 2008, {The Nearest OB Association:
  Scorpius-Centaurus (Sco OB2)}, ed. B.~{Reipurth} (Astronomical Society of the
  Pacific), 235

\bibitem[{{Prusti} {et~al.}(2016){Prusti}, {de Bruijne}, {Brown}, {Vallenari},
  {Babusiaux}, {Bailer-Jones}, {Bastian}, {Biermann}, {Evans}, {Eyer},
  {Jansen}, {Jordi}, {Klioner}, {Lammers}, {Lindegren}, {Luri}, {Mignard},
  {Milligan}, {Panem}, {Poinsignon}, {Pourbaix}, {Randich}, {Sarri},
  {Sartoretti}, {Siddiqui}, {Soubiran}, {Valette}, {van Leeuwen}, {Walton},
  {Aerts}, {Arenou}, {Cropper}, {Drimmel}, {H{\o}g}, {Katz}, {Lattanzi},
  {O'Mullane}, {Grebel}, {Holland}, {Huc}, {Passot}, {Bramante}, {Cacciari},
  {Casta{\~n}eda}, {Chaoul}, {Cheek}, {De Angeli}, {Fabricius}, {Guerra},
  {Hern{\'a}ndez}, {Jean-Antoine-Piccolo}, {Masana}, {Messineo}, {Mowlavi},
  {Nienartowicz}, {Ord{\'o}{\~n}ez-Blanco}, {Panuzzo}, {Portell}, {Richards},
  {Riello}, {Seabroke}, {Tanga}, {Th{\'e}venin}, {Torra}, {Els}, {Gracia-
  Abril}, {Comoretto}, {Garcia-Reinaldos}, {Lock}, {Mercier}, {Altmann},
  {Andrae}, {Astraatmadja}, {Bellas-Velidis}, {Benson}, {Berthier}, {Blomme},
  {Busso}, {Carry}, {Cellino}, {Clementini}, {Cowell}, {Creevey}, {Cuypers},
  {Davidson}, {De Ridder}, {de Torres}, {Delchambre}, {Dell'Oro}, {Ducourant},
  {Fr{\'e}mat}, {Garc{\'\i}a-Torres}, {Gosset}, {Halbwachs}, {Hambly},
  {Harrison}, {Hauser}, {Hestroffer}, {Hodgkin}, {Huckle}, {Hutton},
  {Jasniewicz}, {Jordan}, {Kontizas}, {Korn}, {Lanzafame}, {Manteiga},
  {Moitinho}, {Muinonen}, {Osinde}, {Pancino}, {Pauwels}, {Petit},
  {Recio-Blanco}, {Robin}, {Sarro}, {Siopis}, {Smith}, {Smith}, {Sozzetti},
  {Thuillot}, {van Reeven}, {Viala}, {Abbas}, {Abreu Aramburu}, {Accart},
  {Aguado}, {Allan}, {Allasia}, {Altavilla}, {{\'A}lvarez}, {Alves},
  {Anderson}, {Andrei}, {Anglada Varela}, {Antiche}, {Antoja}, {Ant{\'o}n},
  {Arcay}, {Atzei}, {Ayache}, {Bach}, {Baker}, {Balaguer-N{\'u}{\~n}ez},
  {Barache}, {Barata}, {Barbier}, {Barblan}, {Baroni}, {Barrado y
  Navascu{\'e}s}, {Barros}, {Barstow}, {Becciani}, {Bellazzini}, {Bellei},
  {Bello Garc{\'\i}a}, {Belokurov}, {Bendjoya}, {Berihuete}, {Bianchi},
  {Bienaym{\'e}}, {Billebaud}, {Blagorodnova}, {Blanco-Cuaresma}, {Boch},
  {Bombrun}, {Borrachero}, {Bouquillon}, {Bourda}, {Bouy}, {Bragaglia},
  {Breddels}, {Brouillet}, {Br{\"u}semeister}, {Bucciarelli}, {Budnik},
  {Burgess}, {Burgon}, {Burlacu}, {Busonero}, {Buzzi}, {Caffau}, {Cambras},
  {Campbell}, {Cancelliere}, {Cantat-Gaudin}, {Carlucci}, {Carrasco},
  {Castellani}, {Charlot}, {Charnas}, {Charvet}, {Chassat}, {Chiavassa},
  {Clotet}, {Cocozza}, {Collins}, {Collins}, {Costigan}, {Crifo}, {Cross},
  {Crosta}, {Crowley}, {Dafonte}, {Damerdji}, {Dapergolas}, {David}, {David},
  {De Cat}, {de Felice}, {de Laverny}, {De Luise}, {De March}, {de Martino},
  {de Souza}, {Debosscher}, {del Pozo}, {Delbo}, {Delgado}, {Delgado}, {di
  Marco}, {Di Matteo}, {Diakite}, {Distefano}, {Dolding}, {Dos Anjos},
  {Drazinos}, {Dur{\'a}n}, {Dzigan}, {Ecale}, {Edvardsson}, {Enke}, {Erdmann},
  {Escolar}, {Espina}, {Evans}, {Eynard Bontemps}, {Fabre}, {Fabrizio},
  {Faigler}, {Falc{\~a}o}, {Farr{\`a}s Casas}, {Faye}, {Federici}, {Fedorets},
  {Fern{\'a}ndez-Hern{\'a}ndez}, {Fernique}, {Fienga}, {Figueras}, {Filippi},
  {Findeisen}, {Fonti}, {Fouesneau}, {Fraile}, {Fraser}, {Fuchs}, {Furnell},
  {Gai}, {Galleti}, {Galluccio}, {Garabato}, {Garc{\'\i}a-Sedano}, {Gar{\'e}},
  {Garofalo}, {Garralda}, {Gavras}, {Gerssen}, {Geyer}, {Gilmore}, {Girona},
  {Giuffrida}, {Gomes}, {Gonz{\'a}lez-Marcos}, {Gonz{\'a}lez-N{\'u}{\~n}ez},
  {Gonz{\'a}lez-Vidal}, {Granvik}, {Guerrier}, {Guillout}, {Guiraud},
  {G{\'u}rpide}, {Guti{\'e}rrez-S{\'a}nchez}, {Guy}, {Haigron},
  {Hatzidimitriou}, {Haywood}, {Heiter}, {Helmi}, {Hobbs}, {Hofmann}, {Holl},
  {Holland}, {Hunt}, {Hypki}, {Icardi}, {Irwin}, {Jevardat de Fombelle},
  {Jofr{\'e}}, {Jonker}, {Jorissen}, {Julbe}, {Karampelas}, {Kochoska},
  {Kohley}, {Kolenberg}, {Kontizas}, {Koposov}, {Kordopatis}, {Koubsky},
  {Kowalczyk}, {Krone-Martins}, {Kudryashova}, {Kull}, {Bachchan},
  {Lacoste-Seris}, {Lanza}, {Lavigne}, {Le Poncin-Lafitte}, {Lebreton},
  {Lebzelter}, {Leccia}, {Leclerc}, {Lecoeur-Taibi}, {Lemaitre}, {Lenhardt},
  {Leroux}, {Liao}, {Licata}, {Lindstr{\o}m}, {Lister}, {Livanou}, {Lobel},
  {L{\"o}ffler}, {L{\'o}pez}, {Lopez-Lozano}, {Lorenz}, {Loureiro},
  {MacDonald}, {Magalh{\~a}es Fernandes}, {Managau}, {Mann}, {Mantelet},
  {Marchal}, {Marchant}, {Marconi}, {Marie}, {Marinoni}, {Marrese},
  {Marschalk{\'o}}, {Marshall}, {Mart{\'\i}n-Fleitas}, {Martino}, {Mary},
  {Matijevi{\v{c}}}, {Mazeh}, {McMillan}, {Messina}, {Mestre}, {Michalik},
  {Millar}, {Miranda}, {Molina}, {Molinaro}, {Molinaro}, {Moln{\'a}r},
  {Moniez}, {Montegriffo}, {Monteiro}, {Mor}, {Mora}, {Morbidelli}, {Morel},
  {Morgenthaler}, {Morley}, {Morris}, {Mulone}, {Muraveva}, {Musella},
  {Narbonne}, {Nelemans}, {Nicastro}, {Noval}, {Ord{\'e}novic},
  {Ordieres-Mer{\'e}}, {Osborne}, {Pagani}, {Pagano}, {Pailler}, {Palacin},
  {Palaversa}, {Parsons}, {Paulsen}, {Pecoraro}, {Pedrosa}, {Pentik{\"a}inen},
  {Pereira}, {Pichon}, {Piersimoni}, {Pineau}, {Plachy}, {Plum}, {Poujoulet},
  {Pr{\v{s}}a}, {Pulone}, {Ragaini}, {Rago}, {Rambaux}, {Ramos-Lerate},
  {Ranalli}, {Rauw}, {Read}, {Regibo}, {Renk}, {Reyl{\'e}}, {Ribeiro},
  {Rimoldini}, {Ripepi}, {Riva}, {Rixon}, {Roelens}, {Romero-G{\'o}mez},
  {Rowell}, {Royer}, {Rudolph}, {Ruiz-Dern}, {Sadowski}, {Sagrist{\`a}
  Sell{\'e}s}, {Sahlmann}, {Salgado}, {Salguero}, {Sarasso}, {Savietto},
  {Schnorhk}, {Schultheis}, {Sciacca}, {Segol}, {Segovia}, {Segransan},
  {Serpell}, {Shih}, {Smareglia}, {Smart}, {Smith}, {Solano}, {Solitro},
  {Sordo}, {Soria Nieto}, {Souchay}, {Spagna}, {Spoto}, {Stampa}, {Steele},
  {Steidelm{\"u}ller}, {Stephenson}, {Stoev}, {Suess}, {S{\"u}veges}, {Surdej},
  {Szabados}, {Szegedi-Elek}, {Tapiador}, {Taris}, {Tauran}, {Taylor},
  {Teixeira}, {Terrett}, {Tingley}, {Trager}, {Turon}, {Ulla}, {Utrilla},
  {Valentini}, {van Elteren}, {Van Hemelryck}, {van Leeuwen}, {Varadi},
  {Vecchiato}, {Veljanoski}, {Via}, {Vicente}, {Vogt}, {Voss}, {Votruba},
  {Voutsinas}, {Walmsley}, {Weiler}, {Weingrill}, {Werner}, {Wevers},
  {Whitehead}, {Wyrzykowski}, {Yoldas}, {{\v{Z}}erjal}, {Zucker}, {Zurbach},
  {Zwitter}, {Alecu}, {Allen}, {Allende Prieto}, {Amorim},
  {Anglada-Escud{\'e}}, {Arsenijevic}, {Azaz}, {Balm}, {Beck}, {Bernstein},
  {Bigot}, {Bijaoui}, {Blasco}, {Bonfigli}, {Bono}, {Boudreault}, {Bressan},
  {Brown}, {Brunet}, {Bunclark}, {Buonanno}, {Butkevich}, {Carret}, {Carrion},
  {Chemin}, {Ch{\'e}reau}, {Corcione}, {Darmigny}, {de Boer}, {de Teodoro}, {de
  Zeeuw}, {Delle Luche}, {Domingues}, {Dubath}, {Fodor}, {Fr{\'e}zouls},
  {Fries}, {Fustes}, {Fyfe}, {Gallardo}, {Gallegos}, {Gardiol}, {Gebran},
  {Gomboc}, {G{\'o}mez}, {Grux}, {Gueguen}, {Heyrovsky}, {Hoar}, {Iannicola},
  {Isasi Parache}, {Janotto}, {Joliet}, {Jonckheere}, {Keil}, {Kim},
  {Klagyivik}, {Klar}, {Knude}, {Kochukhov}, {Kolka}, {Kos}, {Kutka}, {Lainey},
  {LeBouquin}, {Liu}, {Loreggia}, {Makarov}, {Marseille}, {Martayan},
  {Martinez-Rubi}, {Massart}, {Meynadier}, {Mignot}, {Munari}, {Nguyen},
  {Nordlander}, {Ocvirk}, {O'Flaherty}, {Olias Sanz}, {Ortiz}, {Osorio},
  {Oszkiewicz}, {Ouzounis}, {Palmer}, {Park}, {Pasquato}, {Peltzer}, {Peralta},
  {P{\'e}turaud}, {Pieniluoma}, {Pigozzi}, {Poels}, {Prat}, {Prod'homme},
  {Raison}, {Rebordao}, {Risquez}, {Rocca-Volmerange}, {Rosen}, {Ruiz-Fuertes},
  {Russo}, {Sembay}, {Serraller Vizcaino}, {Short}, {Siebert}, {Silva},
  {Sinachopoulos}, {Slezak}, {Soffel}, {Sosnowska}, {Strai{\v{z}}ys}, {ter
  Linden}, {Terrell}, {Theil}, {Tiede}, {Troisi}, {Tsalmantza}, {Tur},
  {Vaccari}, {Vachier}, {Valles}, {Van Hamme}, {Veltz}, {Virtanen}, {Wallut},
  {Wichmann}, {Wilkinson}, {Ziaeepour}, \& {Zschocke}}]{gaia_mission}
{Prusti}, T., {de Bruijne}, J.~H.~J., {Brown}, A.~G.~A., {et~al.} 2016, \aap,
  595, A1

\bibitem[{{Reino} {et~al.}(2018){Reino}, {de Bruijne}, {Zari}, {d'Antona}, \&
  {Ventura}}]{Reino18}
{Reino}, S., {de Bruijne}, J., {Zari}, E., {d'Antona}, F., \& {Ventura}, P.
  2018, \mnras, 477, 3197

\bibitem[{{Riedel} {et~al.}(2017){Riedel}, {Blunt}, {Lambrides}, {Rice},
  {Cruz}, \& {Faherty}}]{riedel17}
{Riedel}, A.~R., {Blunt}, S.~C., {Lambrides}, E.~L., {et~al.} 2017, \aj, 153,
  95

\bibitem[{{R{\"o}ser} {et~al.}(2018){R{\"o}ser}, {Schilbach}, \&
  {Goldman}}]{roeser18}
{R{\"o}ser}, S., {Schilbach}, E., \& {Goldman}, B. 2018, ArXiv e-prints,
  arXiv:1811.03845

\bibitem[{{R{\"o}ser} {et~al.}(2011){R{\"o}ser}, {Schilbach}, {Piskunov},
  {Kharchenko}, \& {Scholz}}]{Roeser11}
{R{\"o}ser}, S., {Schilbach}, E., {Piskunov}, A.~E., {Kharchenko}, N.~V., \&
  {Scholz}, R.~D. 2011, \aap, 531, A92

\bibitem[{{Sch{\"o}nrich} {et~al.}(2010){Sch{\"o}nrich}, {Binney}, \&
  {Dehnen}}]{schoenrich10}
{Sch{\"o}nrich}, R., {Binney}, J., \& {Dehnen}, W. 2010, \mnras, 403, 1829

\bibitem[{{Spitzer}(1940)}]{Spitzer1940}
{Spitzer}, Lyman, J. 1940, \mnras, 100, 396

\bibitem[{{Terlevich}(1987)}]{Terlevich87}
{Terlevich}, E. 1987, \mnras, 224, 193

\bibitem[{{van der Walt} {et~al.}(2011){van der Walt}, {Colbert}, \&
  {Varoquaux}}]{numpy}
{van der Walt}, S., {Colbert}, S.~C., \& {Varoquaux}, G. 2011, Computing in
  Science and Engg., 13, 22

\bibitem[{van~der Walt {et~al.}(2014)van~der Walt, {S}ch\"onberger,
  {Nunez-Iglesias}, {B}oulogne, {W}arner, {Y}ager, {G}ouillart, {Y}u, \& the
  scikit-image contributors}]{scikit-image}
van~der Walt, S., {S}ch\"onberger, J.~L., {Nunez-Iglesias}, J., {et~al.} 2014,
  PeerJ, 2, e453

\bibitem[{{Wenger} {et~al.}(2000){Wenger}, {Ochsenbein}, {Egret}, {Dubois},
  {Bonnarel}, {Borde}, {Genova}, {Jasniewicz}, {Lalo{\"e}}, {Lesteven}, \&
  {Monier}}]{simbad}
{Wenger}, M., {Ochsenbein}, F., {Egret}, D., {et~al.} 2000, \aaps, 143, 9

\end{thebibliography}

\begin{appendix}

\section{Coordinate system definitions}
\label{app:coordinates}

The following list summarizes our parameter setup for the calculation of Galactocentric cylindrical velocities.

\begin{itemize}
    \item $v_R$; radial velocity component. Positive when moving toward the Galactic center.
    \item $v_{\phi}$; tangential velocity component. Positive in the direction of Galactic rotation.
    \item $v_Z$; vertical velocity component. Positive toward the Galactic North Pole.
    \item $r_0$ = \SI{8300}{pc}; distance from vantage point to the Galactic center \citep{Gillessen09}.
    \item $\phi_0$ = \SI{180}{deg}; position angle of the Sun in Galactocentric cylindrical coordinates.
    \item $z_0$ = \SI{27}{pc}; height of the Sun above the Galactic plane toward the Galactic North Pole \citep{Chen01}.
    \item $v_0$ = \SI{220}{\km \per \s}; circular velocity at $r_0$ \citep{Kerr86}.
    \item $(U, V, W)_{\mathrm{solar}}$ = (11.1, 12.24, 7.25) \SI{}{\km \per \s}; barycentric velocity of the Sun relative to the local standard of rest (\citealp{schoenrich10}).
\end{itemize}

\section{Completeness}
\label{sec:app:completeness}

\begin{figure}[ht!]
        \centering
        \resizebox{\hsize}{!}{\includegraphics[]{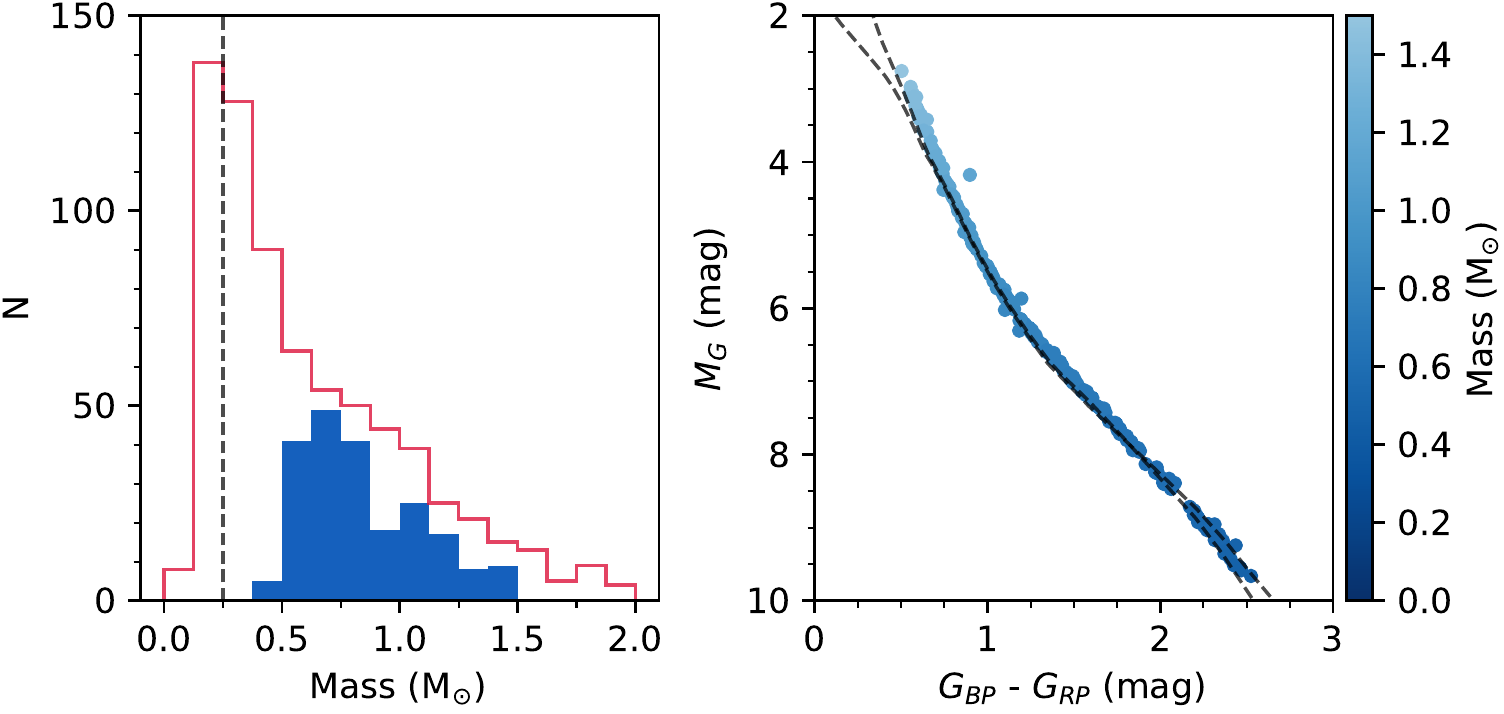}}
        \caption[]{Mass spectrum of our Hyades member selection. The left panel shows a histogram of the derived masses, where the blue bars show our selection. The red histogram was constructed with data from \citet[][]{Roeser11}. The panel to the right displays an observational HRD with data color-coded by mass. The isochrones are the same as in Fig.\ref{img:hrd}.}
    \label{img:mass_function}
\end{figure}

To investigate the completeness of our sample, we derived the mass spectrum for our selection by applying a nearest-neighbor interpolation on the \SI{1}{Gyr}, $Z=0.02$ PARSEC isochrone. The full mass spectrum of our selection is shown in Fig.~\ref{img:mass_function} and covers a range from about \SI{0.5}{M_\odot} to \SI{1.5}{M_\odot}. In this figure, we also show the mass spectrum for the data published by \citet[][]{Roeser11}. The authors quote an approximate sensitivity limit of about \SI{0.25}{M_\odot} , which is marked as a vertical dashed line. Comparing the masses derived by \citet[][]{Roeser11} and
our isochrone interpolation for matching sources, we find a difference of $0.0 \pm 0.18$ \SI{}{M_\odot} and therefore no systematic bias. The mass spectrum in the left plot in Fig.~\ref{img:mass_function} clearly shows that our selection suffers from incompleteness across all mass ranges, where we both miss the most massive members, as well as stars with masses below $\sim$\SI{0.5}{M_\odot}.

\section{All-sky distribution}
\label{sec:app:allsky}

\begin{figure}[ht!]
        \centering
        \resizebox{\hsize}{!}{\includegraphics[]{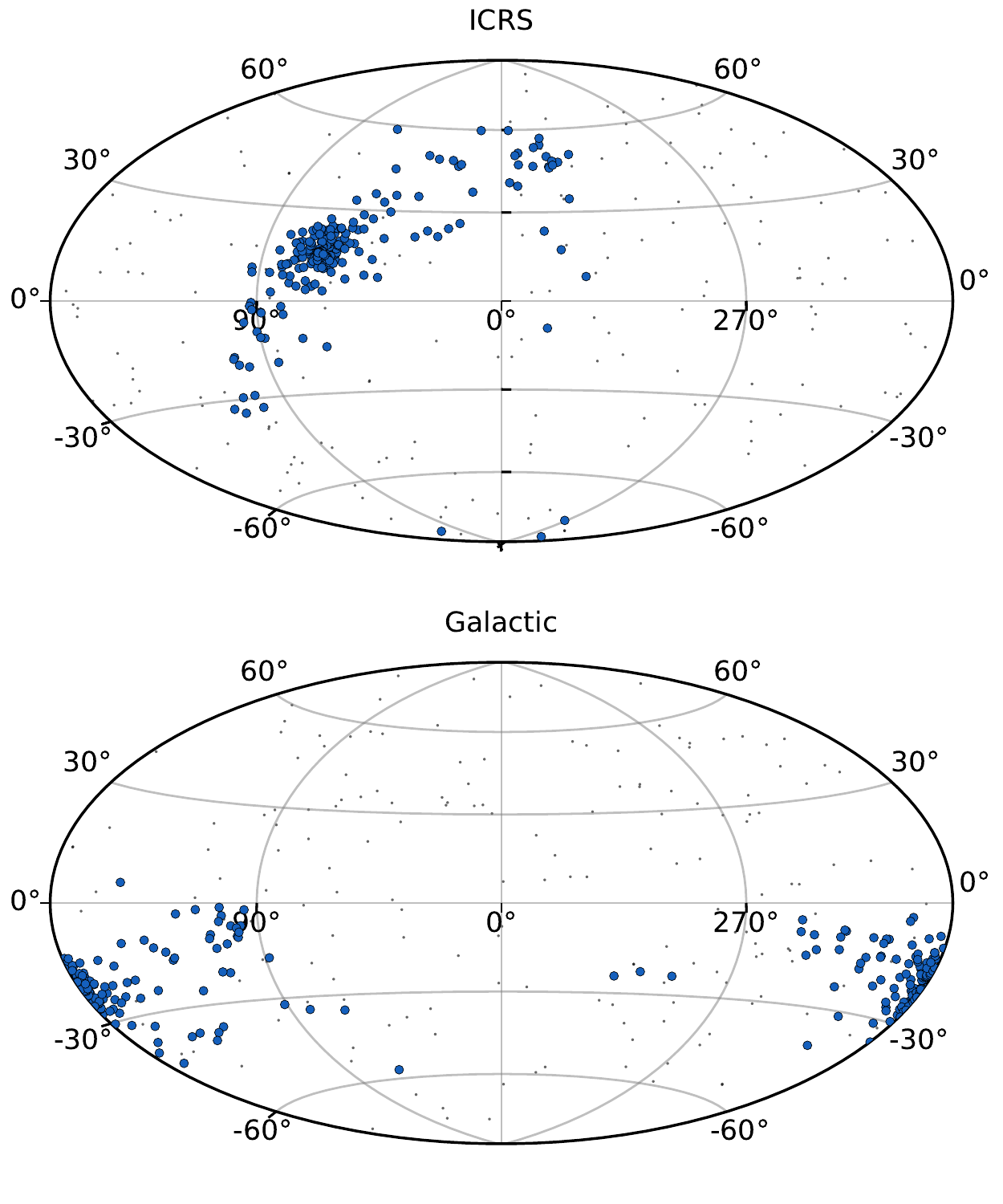}}
        \caption[]{Source positions for Hyades cluster members as determined in this letter shown in Equatorial (top) and Galactic (bottom) coordinates. The blue points are our final member selection, the gray dots in the background refer to all sources. In total, the entire structure spans a size of more than 100~degrees on the sky.}
    \label{img:allsky}
\end{figure}

Figure~\ref{img:allsky} shows our Hyades selection as projected onto the sky. The top panel shows Equatorial coordinates, and the bottom panel presents Galactic coordinates.

\end{appendix}

\end{document}